\documentclass[aps,twocolumn,letterpaper,prl,showpacs, 10pt]{revtex4-1}
\usepackage{amsmath}
\usepackage{eurosym}
\usepackage{graphicx}
\usepackage{dcolumn}
\usepackage{bm}
\usepackage{color}
\usepackage{latexsym}
\usepackage{amssymb}
\usepackage{subfigure}
\begin{document}

\title{Geometry for a `penguin-albatross' rookery.}
\author{Fabio Giavazzi$^{(1)}$}
\email[]{fabio.giavazzi@unimi.it}
\author{Alberto Vailati$^{(2)}$}
\email[]{alberto.vailati@unimi.it}
\affiliation{$^{(1)}$Dipartimento, di Biotecnologie Mediche e Medicina Traslazionale, Universit\`a degli Studi di Milano, Milano, Italy}
\affiliation{$^{(2)}$ Dipartimento di Fisica, Universit\`a degli Studi di Milano, I-20133 Milano, Italy}
\date{\today}

\begin{abstract}
We introduce a simple ecological model describing the spatial organization of two interacting
populations whose individuals are indifferent to conspecifics and avoid the proximity to heterospecifics. At
small population densities $\Phi$ a non-trivial structure is observed where clusters of individuals arrange into a rhomboidal bipartite network with an average degree of four.
For $\Phi\rightarrow0$ the length scale, order parameter and susceptibility of the network exhibit power-law divergences compatible with hyper-scaling, suggesting the existence of a zero density - non-trivial - critical point.
At larger densities a critical threshold  $\Phi_{c}$ is identified above which the evolution toward a partially ordered configuration is  prevented and the system becomes jammed in a fully mixed state.

%Valid PACS numbers may be entered using the \verb+\pacs{#1}+ command.
\end{abstract}

%\keywords{population dynamics, pattern formation, bipartite network, critical phenomena, percolation}
\maketitle

%   \cite{}

\bigskip

\section{I. INTRODUCTION}
The interactions within and between species, either for survival or for the control of
the territory, gives rise to a rich spatiotemporal dynamics \cite{couzin2003, vicsek2012}, the most investigated case being probably the two species
prey-predator system \cite{vasseur2009, kamimura2010, vicsek2010, angelani2012}. Models
and experiments have outlined the emergence of collective behaviour and pattern formation arising
from simple local rules, a prototypical example being represented by the selfish herd model \cite{hamilton1971, king2012}.
The study of the complex interplay between inter- and intraspecific interactions and the emergence, at different scales, of structures in the spatial distribution of the individuals, is an important topic in population ecology \cite{legendre1989}. Within this framework, the investigation of systems incorporating species competing symmetrically, or quasi-symmetrically, for space has attracted interest recently when it has been shown that it leads to the development of a robust diversification of ecosystems \cite{mathiesen2011,mitarai2012, sneppen2013}. The symmetrical interaction can lead to the spatial segregation of species \cite{gotelli2010}, and a competitive advantage can be determined by the emergent formation of spatial structures \cite{ heinsalu2013}. Even conspecific seabirds exhibit a spatial segregation into non overlapping colonies feeding from mutually exclusive areas \cite{wakefield2013, weimerskirch2013}. Under some circumstances, different species organize into mixed colonies. A nice example is represented by the association between Rockhopper penguin (\textit{Eudyptes chrysocome}) and Black-browed albatross (\textit{Thalassarche melanophrys}) into common colonies, reported by several investigators involved in the census of birds in the Falkland Islands \cite{strange1982, huin2007, demongin2009, crofts2011, strange2011, baylis2012}.
This peculiar system has attracted the attention of several novelists in 19\textsuperscript{th} century \cite{morrell, poe, verne}, who gave detailed - allegedly fictional - descriptions of the non-trivial topological distribution of the individuals of the two species. In particular, B. Morrell in \cite{morrell} described a penguin-albatross rookery
where individuals are uniformly distributed on a nesting area in a staggered
configuration where ``each albatross is surrounded by four penguins; and each
penguin has an albatross for its neighbor, in four directions''. 

In this work we argue that the description of the topology of a penguin-albatross rookery by Morrell and Poe could be based on the empirical observation of a real colony, rather than on imagination. Taking inspiration from them, we investigate numerically the evolution of the spatial distribution of two populations of ``penguins'' and ``albatrosses'' sharing the same area. We show that structures similar to those described in their narratives form by assuming that individuals behave according to two minimal rules: \textit{i}$)$ individuals are indifferent to conspecifics (except for avoiding superposition) and \textit{ii}$)$ avoid the proximity to heterospecifics. These behavioral rules determine a tendency of the system to phase-separate into two clusters made of conspecific individuals. However, during the separation process the system becomes jammed into a structurally
arrested configuration whose spatial structure strongly depends on the population density. At small population densities clusters of individuals arrange into a bipartite network spanning the entire system. The network is entirely made of rhomboidal cells, where two clusters of conspecific individuals sit at two opposite
corners and two clusters of individuals of the other species occupy the other two corners. This partially
ordered phase is characterized by a single dominant length scale represented
by the distance between nearest heterospecific individuals and exhibits a
non-trivial critical point at vanishing population density, characterized by a divergence of the order parameter in the presence of hyperscaling. At larger population densities the system becomes frozen into a structurally arrested configuration, where the motion of individuals is caged by those surrounding it.
%%%%FIG.01
\begin{figure*}[t]
\centering\includegraphics[width=15cm] {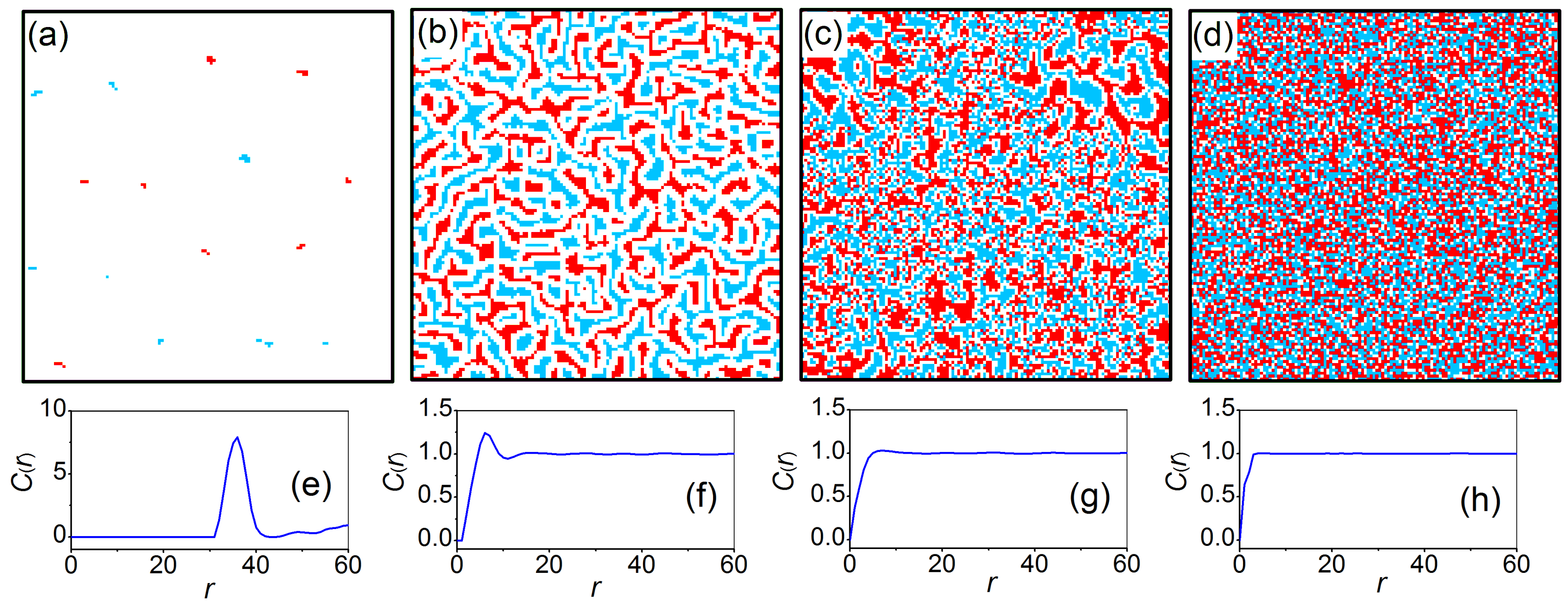}
\caption{(color online) patterns formed at steady state at different population densities. Red/ light blue (dark gray/light gray) pixels mark the position of individuals of the two species, while empty sites are white. (a) $\Phi=0.002$, (b) $\Phi=0.2$, (c) $\Phi=0.31$ and (d) $\Phi=0.4$. Panels (e)-(h) show the corresponding cross-correlation functions. The evolution of the patterns is shown in Supplemental Material movie S1 \cite{supplemental}. The size of the simulation box is 128$\times$128 sites.}
\label{pat}
\end{figure*}

\section{II. MODEL ECOSYSTEM}

Our model is of general relevance, but for simplicity we refer to the populations as ``Penguins'' (P) and ``Albatrosses'' (A). We assume that each population has an equal number $N$ of individuals \cite{ratio}, initially distributed randomly with density $\Phi
=N/M^{2}$ across the sites of a square lattice of side $M=128$ with periodic
boundary conditions.

The system evolves trough discrete time steps. At each time step, an individual is randomly chosen from the two population with uniform probability. Let $d$ be the Euclidean distance between the selected individual and the nearest individual of the opposite species. The individual attempts a move towards a new site, randomly chosen between one its four neighboring lattice sites. If the new site is occupied by another individual, or if the distance $d'$ between the new site and the nearest individual of the opposite species is smaller than $d$, the move is rejected. Otherwise it is accepted, and the position of the selected individual is updated accordingly.

In all the simulations discussed in the present work the total number of time steps $N_s$ is chosen in order to ensure that a steady state is reached. A reliable indicator of the convergence to a statistically stable state is represented by the cross-correlation length $\xi _{{c}}$ (see section III.B.).  During the evolution of the system a monotonic increase of the cross-correlation length  is observed, until it reaches a plateau. For each population density a few exploratory simulations were run and a characteristic equilibration time $n_{0}$ was estimated by fitting $\xi _{c}(n)$ as a function of the simulation step $n$ with a stretched exponential function: $f\left(n\right)=A\left\{ 1-\exp\left[-\left(n/n_{0}\right)^{\delta}\right]\right\} +B$. The duration $N_s$ was then set to a value at least ten time larger than $n_{0}$.

 \section{III. RESULTS}

 The spatiotemporal dynamics of the system is strongly affected by the population density. For small $\Phi $, individuals of one species form small clusters surrounded on average by four clusters of the other species (Fig. 1(a), Supplemental Material movie \cite{supplemental}), a configuration partially reminiscent of that depicted by the narratives of the 19\textsuperscript{th} century. We define a cluster as a connected set of individuals of the same species, say Penguins: two penguins (sitting on sites $x$ and $x^\prime$, respectively) belong to the same cluster if a continuous path exists connecting $x$ and $x^\prime$ made up of steps connecting neighboring sites all occupied by penguins.
 At intermediate $\Phi $ individuals arrange into large connected fractal
clusters staggered across the plane [Fig. 1(b), Supplemental Material movie \cite{supplemental}]. These
clusters are separated by convoluted aisles free of individuals. At a
critical density $\Phi _{{c}}\approx $0.311 the system undergoes a
transition to a disordered phase [Fig. 1(c), Supplemental Material movie \cite{supplemental}]. The free
aisles disappear and the distribution of species exhibits large
concentration fluctuations determined by the strong spatial heterogeneity of
the configurations (see the late stages of Supplemental Material movie \cite{supplemental}). At $\Phi >\Phi _{c}$ the system becomes jammed into a disordered
configuration [Fig. 1(d), Supplemental Material movie \cite{supplemental}] where the limited number of
free sites determines a structural arrest that prevents the development of
long range correlations, similar to what happens in a glassy state.  Thus the system exhibits two critical points at $\Phi_{0}=0 $ and $\Phi_{c}=0.311 $ In the following we will describe the properties of these two critical points separately.

\subsection{A. Structural arrest at high $\Phi$ }

Due to
the inter-specific repulsion between individuals, a structural arrest transition takes place
at packing fractions well below the close packing of individuals needed for
the occurrence of the huddling transition reported in a single species \cite{zitterbart2011,canals2011, zitterbart2013}. The transition is characterized by the
appearance of clusters of individuals with fractal dimension $d_{f} \approx 1.4$.
The cluster size distribution close to the critical point is compatible with a power-law with exponent -1.4, with an exponential cutoff. By
choosing the average cluster size $\Psi _{a}$ as a suitable order
parameter one can appreciate that the transition is characterized by a
continuous decrease of the order parameter [Fig. 2(a)], accompanied by a
divergence of the susceptibility of the system $\chi _{a} $ [Fig. 2(b)],
which represents the variance of the order parameter across different realizations. The average correlation
length $\xi _{{a}}$ of the clusters increases continuously at the
transition [Fig. 2(c)], without showing any sign of divergence. This feature
suggests that the critical behavior is not determined by the development of
intra-species long range order.

%%%%FIG.02
\begin{figure}[t]
\centering\includegraphics[width=7cm]{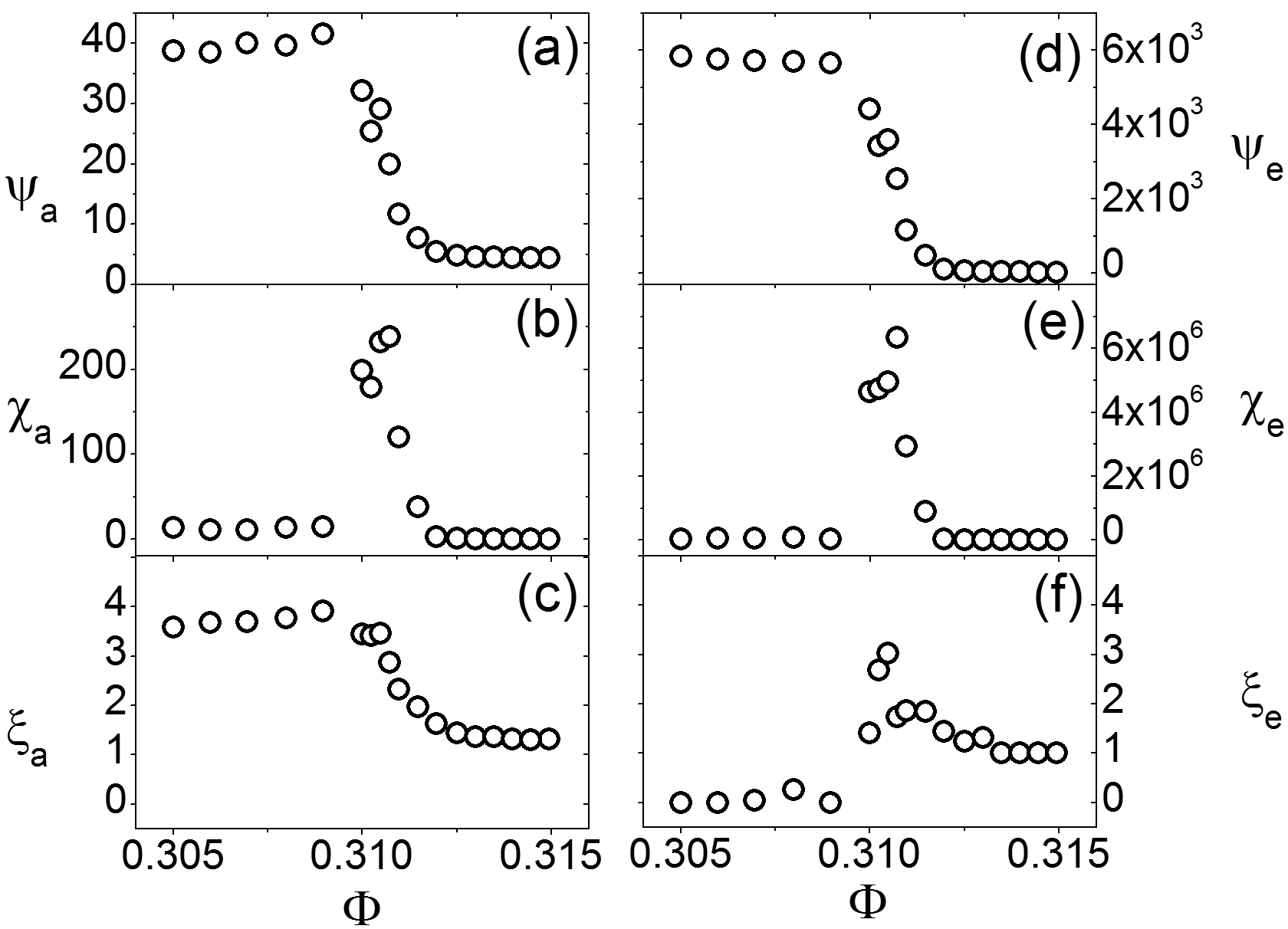}
\caption{(a)-(d) order parameter, (b)-(e) susceptibility and (c)-(f) correlation length of the system. The left column corresponds to a single species, while the right one to the empty region.}
\label{crit}
\end{figure}

The critical transition can be best understood
as due to a percolation \cite{binder1997} of the empty phase. The availability
of aisles of empty sites spanning the whole lattice represents an essential
ingredient to achieve the long range order exhibited by the system at $\Phi <\Phi _{c}$. Individuals cannot move when
they are stuck inside a cluster of individuals of the same species. The only
region accessible for movement is determined by the empty sites that
separate clusters of different species. In the range 0.1$<\Phi <$0.311 the empty
phase has a labyrinthine structure spanning all the system. The relevant  parameters in this case are the largest cluster size $%
\Psi _{e}$ [Fig. 2(d)], its variance $\chi _{e}$ [Fig. 2(e)] and the
correlation length $\xi _{e}$ of the empty region [Fig. 2(f)],
determined from the size of the second largest cluster of empty sites \cite{binder1997}. The correlation
length $\xi _{e}$ and the susceptibility $\chi _{e}$ diverge close
to the critical point, while the order parameter decreases continuously from
a finite value to zero. This finite value is close to the maximum
cluster size that one would attain by using all the available empty sites. Ideally we would like to determine the critical exponents at the phase transition by a systematic investigation of the finite size scaling  \cite{binder1997}, in order to determine the universality class of the transition. Unfortunately,  the slowing
down of the system dynamics close to the transition (see
Supplemental Material movie \cite{supplemental}) induces divergence of the time needed to reach
a stable configuration, which prevented us from achieving a reliable determination of the exponents.

\subsection{B. Non-trivial critical point at $\Phi=0$ }

At $\Phi <0.31$ the individuals are
able to arrange into clusters and an order related to the mutual spatial
distribution of clusters of different species develops. The position of penguins and albatrosses
is specified by two spatial distributions $P(x)$ and $A(x)$ respectively,
where $P(x)=1$ ($A(x)=1$) for each site occupied by a penguin (albatross)
and 0 otherwise. The statistical relation between the mutual positions of
individuals of the two species is expressed by the radial cross correlation
function $C_{{c}}(r) = \langle P(x)A(x+r) \rangle /\Phi^{2}$. At small $\Phi $ the cross
correlation function exhibits a narrow peak [Fig. 1(e)], while the
autocorrelation functions of $P(x)$ and $A(x)$ are almost featureless (not shown). By increasing $\Phi$ the
cross correlation peak becomes less defined [Fig. 1(f)-(g)] and for $\Phi >
\Phi _{c}$ it disappears completely [Fig. 1(h)], thus marking the transition to a
featureless phase where the mutual position between individuals of different
species is uncorrelated. The position of the peak of $C(r)$
represents the cross-correlation length $\xi _{c}$ of the system [Fig.
3(a)], which provides a typical length scale for the distance between clusters of different species. The contrast of the first peak of the cross
correlation function $\Psi _{c}${\ }= $C_{\max }-1$ represents an 
order parameter suitable to characterize the formation of inter-species ordered structures [Fig. 3(b))]. The variance of the order parameter [Fig. 3(c)] provides the susceptibility $\chi _{c}$ to fluctuations. At small
$\Phi $ these parameters exhibit a power law behavior, $\Psi _{c}\propto $
$\Phi ^{\beta }$, $\chi _{c}\propto $ $\Phi ^{-\gamma }$ and $\xi _{_{c}}$
$\propto $ $\Phi ^{-\nu }$, with exponents $\beta $%
=-0.575$\pm 0.015$, $\gamma $=2.05$\pm 0.06$, and $\nu $=0.37$\pm 0.01$. The exponents are compatible
with the hyperscaling relation $2\beta +\gamma =d\nu $, where $d=2$ is the
dimensionality of the space. We test the goodness of the hyperscaling of our data by plotting the hyperscaling parameter $%
h=\chi _{c}/(\Psi _{c}\xi _{c})^{2}$ [diamonds on Fig. 3(c)]. In the presence of hyperscaling this parameter must be constant. Hyperscaling
suggests that $\Phi _{c}$=0 exhibits the features of a critical point.
%%%%%%FIG.03
\begin{figure}[t]
\centering\includegraphics[width=6cm]{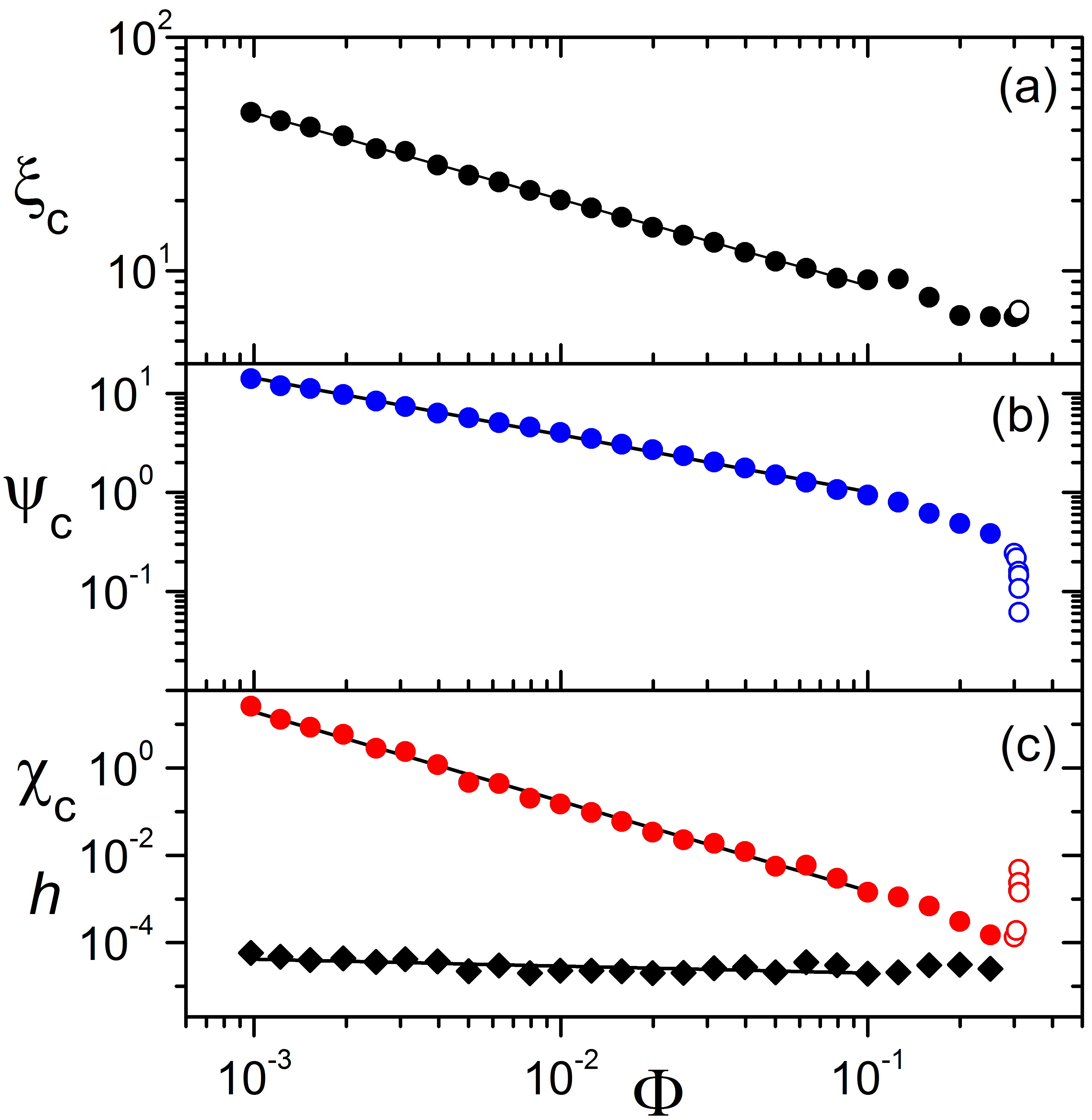}
\caption{(color online) (a) correlation length; (b) order parameter; (c) susceptibility (circles) and hyperscaling parameter (diamonds) characterizing the mutual order of the two species defined from the cross-correlation function $C_{{c}}(r)$. Open and solid symbols represent averages on 20 and 100 independent realizations of the system, respectively. Lines represent the best fit of the data with a power-law. The hyperscaling parameter $h$ (diamonds) is reasonably constant, indicating that the power law exponents are compatible with hyperscaling.}
\label{hyp}
\end{figure}
At variance with the usual behavior of second order phase
transitions, in this case the order parameter diverges close to this
critical point and the critical exponent $\beta $ is negative, a feature
 exhibited by other systems where hyperscaling holds, such
as the thermal denaturation of DNA\cite{buyukdagli2006}. The dynamic evolution of
the patterns at small $\Phi$ is characterized by the transition from a
disordered to an ordered configuration exhibiting on average four-fold
symmetry (Supplemental Material movie \cite{supplemental}). 

The initial configuration of the system
corresponds to a random distribution of individuals of the two species. At
this stage, the domain of influence of each individual is a Voronoi cell
with an average of six sides. During the evolution, individuals belonging to
the same species form clusters. At steady state the spatial distribution of
the clusters is characterized by a bipartite random network \cite{newman2001}
obtained by joining clusters of penguins with the nearest clusters of albatrosses. The network is made of rhomboidal loops where two clusters of penguins sit at opposite
vertices of a rhombus and two cluster of albatrosses at the other two vertices [Fig. 4(a)]. Similar configurations have been reported for the
bioconvection of bacteria \cite{platt1961, janosi1998}, for the spoke pattern
convection of simple fluids occurring at high Rayleigh numbers \cite{busse1974, busse1994} and for the solutal convection of nanoparticles at high
solutal Rayleigh numbers \cite{cerbino2002, mazzoni2008}.
%%%%%FIG.04
\begin{figure}[t]
\centering\includegraphics[width=8.5cm]{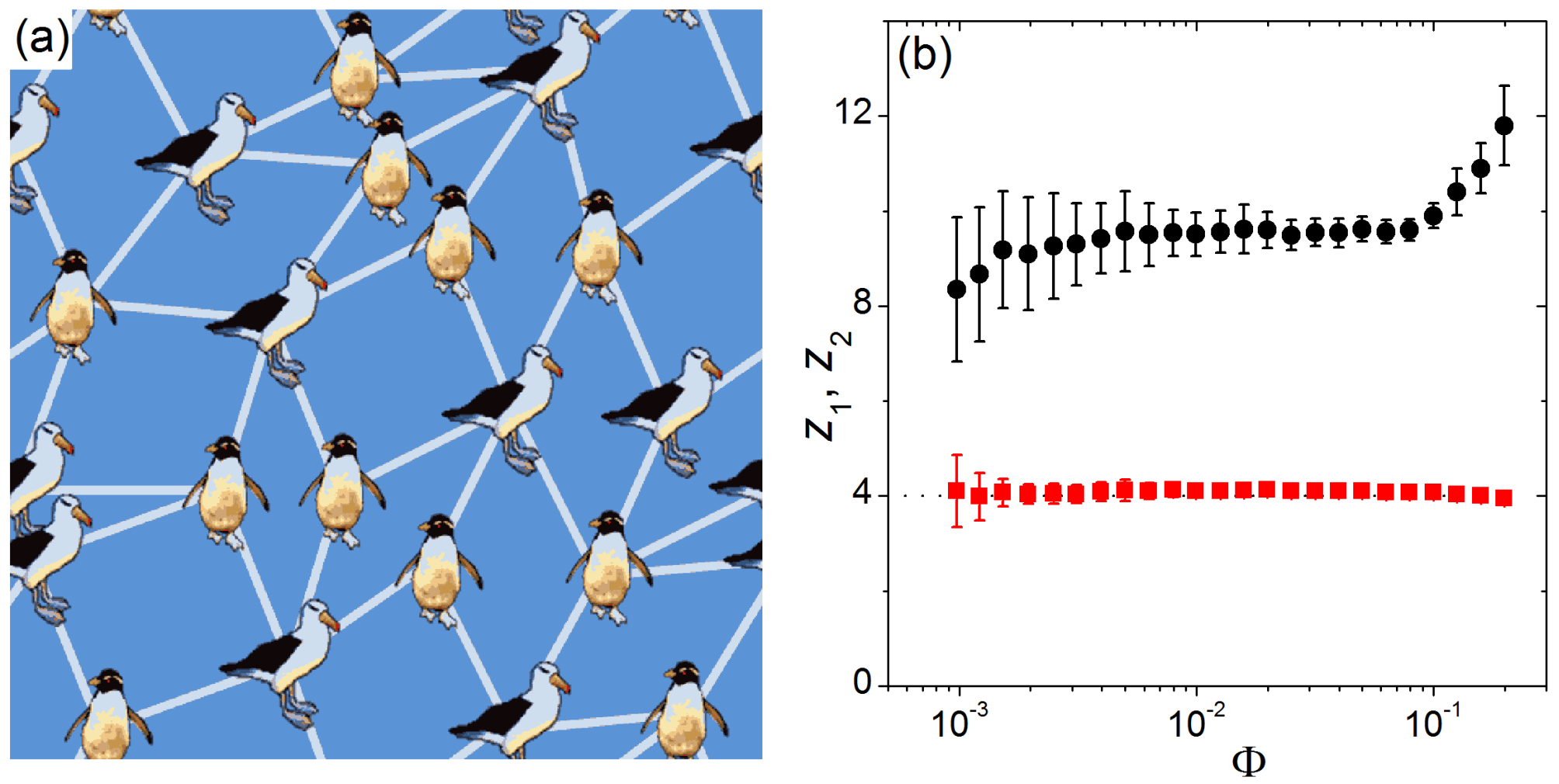}
\caption{(color online) (a) bipartite rhomboidal penguin-albatross network formed at $\Phi=0.002$. (b) average number of first neighbors $z_{1}$ (squares) and of second neighbors (circles) $z_{2}$ of the network. $z_{1}$ remains stable at 4 in a wide range of $\Phi$, in agreement with the narratives by Morrell and Poe. }
\label{net}
\end{figure}

The average
degree $z_{1}$ of a cluster can be determined by averaging the number of
nearest neighbors of each cluster in the network  [Fig. 4(b)] \cite{nearest}. At small $\Phi ,$ $z_{1}\approx 4$, and the
average number of second neighbors is $z_{2}\approx 8$, in agreement with
the fact that on average $z_{1}$ second neighbors are shared by the nearest
neighbors. At larger fractions, $z_{1}$ remains stable at 4, while $z_{2}$
increases to 12, a value compatible with a tree structure of the bipartite
network.

The peculiar pattern exhibited at vanishing $\Phi$ has proven to be robust against perturbations breaking the perfect symmetry between the two species. In particular, we ran simulations in the  presence \textit{i}$)$ of a moderate unbalance in the number $N_1$ and $N_2$ of individuals within each population and \textit{ii}$)$ of a difference in the mobility of the two species, implemented by introducing a systematic bias in the probability $P$ of selecting an individual from a given species for attempting a move. Within the investigated range $ 1\leq\frac{N_{1}}{N_{2}}\leq2$, $0.1\leq P\leq0.5$ no significant variation can be detected in the spatial distributions at steady state when compared to the unperturbed, fully symmetric, situation.

\section{IV. DISCUSSION}

\subsection{A. Topology of the network}

The average four-fold coordination of the vertices of the network can be
understood from Euler's theorem. The theorem states that for an infinite
network $R+V-E=0$, where $R$ is the number of rhomboidal loops, $V$ the number of vertices and $E$ the number of edges connecting
two clusters of different species. Each loop has 4 edges and, since each
edge is shared by two cells, $E=2R$. On the same argument, the number of
edges is related to the number of vertices by $E=z_{1}V/2$. By combining these relations one gets that $z_{1}=4$. Therefore, when the population
density is small, each cluster of
one species is surrounded on average by four clusters of the other
species, a configuration that mirrors the structure reported for the distribution of individuals in the penguin-albatross
rookery described in the narratives by Morrell and Poe \cite{clustsize}. Under this
condition the system is dominated by a single characteristic length scale
corresponding to the average distance between nearest clusters of
different species $\xi _{c}$ [Fig. 3(a)]. This four-fold coordination of the network formed by the two species differs dramatically from the trivial average six-fold coordination, which is expected in colonies of individuals belonging to the same species. Indeed, the development of a four-fold coordination puts more stringent constraints on the mutual position of the individuals than the six-fold coordination, which is also present for a random distribution of individuals \cite{okabe, mazzoni2008}.

\subsection{B. Topology of a penguin-albatross rookery in the narratives of the 19\textsuperscript{th} century}

At the moment, no systematic investigation of the topology of mixed colonies of penguins and albatrosses based on empirical data has been performed to our knowledge. However, the similarities between  the networks generated by our model and those described by Morrell and Poe suggests that their description of a penguin-albatross rookery could be inspired by reality more than expected. The book ``A Narrative of Four Voyages'' by Benjamin Morrell was first published in 1832 by J\&J Harper in New York, while ``The Narrative of Arthur Gordon Pym of Nantucket'' by Edgar Allan Poe was published by the same publisher in 1838. Literary criticism \cite{gitelman} has shown that Poe largely borrowed from Morrell, making it unlikely that Poe himself witnessed real penguin-albatross rookery. Benjamin Morrell was a sea captain and he actually performed several expeditions across the World. The reliability of his narratives has been questioned by historians and geographers, but the fact that his book contains a large amount of accurate information convinced other investigators that his accounts were partially based on reality \cite{gould}. As far as the observation of a penguin albatross-rookery is concerned, Morrell provides a precise description of the geographical location of the colony in New Island, one of the Falkland Islands. Indeed, associations between Rockhopper penguin and Black Browed albatross have regularly been reported from recent bird population censuses on these islands \cite{strange1982, huin2007, demongin2009, crofts2011, strange2011, baylis2012}. This fact, together with the description of the topology of the rookery akin to that described by our model, strongly suggests that the report on the mixed rookery by Morrell could be based on actual observation, rather than on fiction/literary imagination.

\subsection{C. Preliminary comparison with a real mixed colony}

The ideal four-fold coordination reported by Morrell and found in our model system can be difficult to observe in natural systems, due to presence of asymmetries in the behavior or in the interaction between the two species.
As an example, in the case of the penguin-albatross mixed colony, asymmetries can arise from slight differences in the breeding periods and by the constraints in the distributions of individuals determined by peculiarities of the terrain \cite{saino}. In general, one could expect that one of the species settles on the terrain at first, thus providing a template for the final structure of the colony, followed by the second \cite{saino}.
An analysis of real mixed colonies to develop a model that accurately describes all the peculiarities related to temporal and spatial asymmetries would require an extensive set of data showing the time evolution of the position of birds within the colonies during different breeding seasons and in different nesting areas. At the moment we are not aware of the availability of empirical data showing the dynamics of a mixed colony. Indeed, the acquisition of such data from an aerial position through a breeding season would require a significant technical and economical effort.
However, we are in the position of attempting a preliminary analysis by using  an aerial photograph of a mixed colony kindly made available to us by Iain Strange and Georgina Strange [Fig. 5(a)] \cite{strange2011, parallax}. The photograph was taken under uncontrolled conditions: we don't know whether the albatrosses or the penguins occupied the site first and we are not aware of the stage of breeding for the two species. Moreover, during other breeding seasons the conditions can differ significantly from those reported by the photograph. Notwithstanding the peculiar nature of the conditions of the photograph, we think that a preliminary analysis of the geometry of a penguin-albatross rookery is meaningful to start highlighting some general features of a real natural system.

%%%%FIG.05
\begin{figure}[t]
\centering\includegraphics[width=8.5cm]{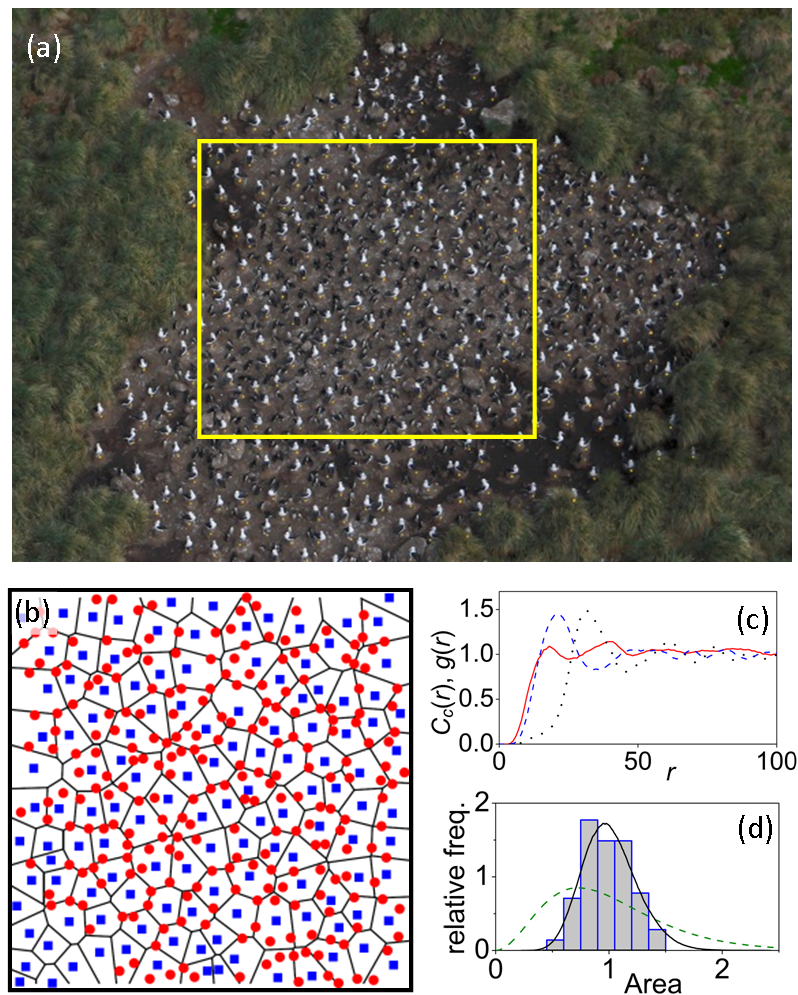}
\caption{(color online) (a) aerial photograph of a Rockhopper penguin - Black-browed albatross colony at Elephant Jason Island, Falkland Islands \cite{strange2011}. Image copyright by Ian J. Strange and Georgina Strange, Design in Nature. The yellow box delimits the region of interest. (b) position of the albatrosses (blue squares) and penguins (red circles) in the region of interest. The network represents the Voronoi diagram generated by albatrosses. The size of the selected region is about 18m x 18m. (c) spatial cross correlation function $C_{{c}}(r)$ for the position of penguins and albatrosses (dotted line); radial distribution functions $g(r)$ of penguins (solid line), and albatrosses (dashed line). (d) histogram of the frequency distribution of the area of the Voronoi cells generated by albatrosses. Areas are normalized by the average area. The solid line represents the Hasegawa-Tanemura distribution \cite{hasegawa1976}. The dashed line corresponds to a random distribution of albatrosses.}
\label{par}
\end{figure}

The presence of an inter-species organization is evidenced by the fact that penguins tend to be located at the boundaries of the Voronoi cells generated by albatrosses [Fig. 5(b)]. Moreover, the cross-correlations function of the position of penguins and albatrosses exhibits a marked peak [Fig. 5(c)]. The radial distribution function of penguins is almost featureless, as it occurs in our model, but that of albatrosses presents a marked peak witnessing the presence of intra-species order [Fig. 5(c)], a feature at variance with our model. Intra-species order is also witnessed by the area distribution of the Voronoi cells generated by albatrosses [Fig. 5(d)]. This distribution is very different from that associated to a set of randomly distributed generators (spatial Poisson process \cite{okabe}), and is compatible with the distribution generated by a Hasegawa-Tanemura adjustment process \cite{hasegawa1976}. In a Hasegawa-Tanemura adjustment process each individual changes its position in the attempt to be as close as possible to the center of its Voronoi cell, defined as the center of mass of the vertices of the cell. In the absence of a second species, such a process leads to the spread across the colony of conspecific individuals. The analysis of the photograph allows us to identify a tentative scenario of the temporal and spatial interactions between the birds that accounts for the observed structures.  1) albatrosses have occupied the nesting site first and spread themselves across the available area. Then they have built their nest. The nests of Black-browed albatrosses are piles made of mud and grass whose position, once established, cannot be changed easily. 2) Once that the nests have been built and the position of albatrosses became stable, penguins have joined and filled the free space. Again, we are not aware whether the penguins shown in Fig. 5(c) had already established the position of their nests. The temporal asymmetry suggested above implies that the species that settled first on the terrain had already reached a stationary condition when the other one joined, thus preventing a mutual redistribution of individuals. As a result of the asymmetric interaction between the species, the structures shown in Fig. 5 are not compatible with the bipartite network predicted by our model and reported by Morrell and Poe.

\section{CONCLUSIONS}

We have shown that a simple model where two species interact by mutually avoiding each other leads to the development of clusters of individuals arranged to form inter-species structures, where each cluster is surrounded on average by four clusters of the other species. The system exhibits a non-trivial critical point at vanishing population densities. At larger population densities, the spatio-temporal dynamics is determined by a percolation of the empty region. A preliminary analysis of empirical data shows that real systems develop inter-species order, but the structure is affected by non-symmetrical interactions between the species.  A
more detailed investigation of a real system with symmetrical interactions is made difficult by the scant empirical data currently available and would ideally require the analysis of time-lapse movies of mixed colonies across several breeding seasons and different breeding sites.

\section{ACKNOWLEDGEMENTS}
We are grateful to G. Zanchetta for bringing to our attention the narrative by E. A. Poe, to I. J. Strange and G. Strange for kindly providing the aerial picture of a mixed colony, to S. Crofts at Falklands Conservation for help with the retrieval of the reports on the census of birds in the Falklands Islands, and to N. Saino for illuminating comments. We thank S. Caracciolo, R.
Cerbino, M. Cosentino Lagomarsino, L. Donati, M. Gherardi, A. Parola, C. La Porta, and S. Zapperi for discussion.


\begin{thebibliography}{99}
\bibitem{nearest} One cluster A is the nearest neighbour of a cluster P of the other species if, and only if, an individual belonging to A is the nearest neighbour of an individual belonging to P.
\bibitem{clustsize} The average number of individuals per cluster decreases as the population density $\Phi$ is decreased. As an example, the average number of individuals per cluster in configurations like that shown in Fig. 1a ($\Phi=0.002$) is of the order of 3, while in the limit of very small $\Phi$ ($\Phi=4E-5$) we find that the average number of individuals per cluster is of the order of 1.5.
\bibitem{parallax}  The image was dilated in the vertical direction by $15\%$ to compensate for the deformation due to a parallax angle of about $30^{\circ}$  with respect to the perpendicular to the terrain. No attempt was made to introduce additional corrections for perspective.
\end{thebibliography}
\end{document}